\documentclass[twoside,twocolumn,9pt]{article}
\usepackage{extsizes}
\usepackage[super,sort&compress,comma]{natbib} 
\usepackage[version=3]{mhchem}
\usepackage[left=1.5cm, right=1.5cm, top=1.785cm, bottom=2.0cm]{geometry}
\usepackage{balance}
\usepackage{times,mathptmx}
\usepackage{sectsty}
\usepackage{graphicx} 
\usepackage{lastpage}
\usepackage[format=plain,justification=justified,singlelinecheck=false,font={stretch=1.125,small,sf},labelfont=bf,labelsep=space]{caption}
\usepackage{float}
\usepackage{fancyhdr}
\usepackage{fnpos}
\usepackage[english]{babel}
\usepackage{array}
\usepackage{droidsans}
\usepackage{charter}
\usepackage[T1]{fontenc}
\usepackage[usenames,dvipsnames]{xcolor}
\usepackage{setspace}
\usepackage[compact]{titlesec}

\usepackage{amsmath,amssymb}
\newcommand{\Alcu}[1]{{\rm Al_2 Cu}}

\definecolor{cream}{RGB}{222,217,201}

\begin{document}

\pagestyle{fancy}
\thispagestyle{plain}
%

\makeFNbottom
\makeatletter
\renewcommand\LARGE{\@setfontsize\LARGE{15pt}{17}}
\renewcommand\Large{\@setfontsize\Large{12pt}{14}}
\renewcommand\large{\@setfontsize\large{10pt}{12}}
\renewcommand\footnotesize{\@setfontsize\footnotesize{7pt}{10}}
\makeatother

\renewcommand{\thefootnote}{\fnsymbol{footnote}}
\renewcommand\footnoterule{\vspace*{1pt}%
\color{cream}\hrule width 3.5in height 0.4pt \color{black}\vspace*{5pt}} 
\setcounter{secnumdepth}{5}

\makeatletter 
\renewcommand\@biblabel[1]{#1}            
\renewcommand\@makefntext[1]%
{\noindent\makebox[0pt][r]{\@thefnmark\,}#1}
\makeatother 
\renewcommand{\figurename}{\small{Fig.}~}
\sectionfont{\sffamily\Large}
\subsectionfont{\normalsize}
\subsubsectionfont{\bf}
\setstretch{1.125} 
\setlength{\skip\footins}{0.8cm}
\setlength{\footnotesep}{0.25cm}
\setlength{\jot}{10pt}
\titlespacing*{\section}{0pt}{4pt}{4pt}
\titlespacing*{\subsection}{0pt}{15pt}{1pt}

\fancyfoot{}
\fancyhead{}
\renewcommand{\headrulewidth}{0pt} 
\renewcommand{\footrulewidth}{0pt}
\setlength{\arrayrulewidth}{1pt}
\setlength{\columnsep}{6.5mm}
\setlength\bibsep{1pt}

\makeatletter 
\newlength{\figrulesep} 
\setlength{\figrulesep}{0.5\textfloatsep} 

\newcommand{\topfigrule}{\vspace*{-1pt}%
\noindent{\color{cream}\rule[-\figrulesep]{\columnwidth}{1.5pt}} }

\newcommand{\botfigrule}{\vspace*{-2pt}%
\noindent{\color{cream}\rule[\figrulesep]{\columnwidth}{1.5pt}} }

\newcommand{\dblfigrule}{\vspace*{-1pt}%
\noindent{\color{cream}\rule[-\figrulesep]{\textwidth}{1.5pt}} }

\makeatother

\twocolumn[
  \begin{@twocolumnfalse}
\vspace{3cm}
\sffamily
\begin{tabular}{m{4.5cm} p{13.5cm} }

& \noindent\LARGE{\textbf{Dynamical solid-liquid transition through oscillatory shear$^\dag$}} \\
\vspace{0.3cm} & \vspace{0.3cm} \\

 & \noindent\large{\'{E}ric Brillaux,$^{\ast}$\textit{$^{a}$} and Francesco Turci\textit{$^{b}$}} \\

 & \noindent\normalsize{Starting from an ideal crystalline state, we numerically study a nonequilibrium dynamical order-disorder transition promoted by the application of a periodic shearing protocol at low temperatures in model systems in two and three dimensions. We observe a continuous (2D) and discontinuous (3D) dynamical transition from an ordered to a disordered steady state. Through the analysis of large-scale simulations, we show that the amorphization mechanism around the discontinuous transition is reminiscent of spinodal decomposition.} \\

\end{tabular}

 \end{@twocolumnfalse} \vspace{0.6cm}

  ]

\renewcommand*\rmdefault{bch}\normalfont\upshape
\rmfamily
\section*{}
\vspace{-1cm}


\footnotetext{\textit{$^{a}$~\'{E}cole Normale Sup\'{e}rieure de Lyon, 65 All\'ee d'Italie, 69007 Lyon, France.}}
\footnotetext{\textit{$^{b}$~H.H. Wills Physics Laboratory, Tyndall Avenue, Bristol, United Kingdom. E-mail: f.turci@bristol.ac.uk}}

\footnotetext{\dag~Electronic Supplementary Information (ESI) available: [details of any supplementary information available should be included here]. See DOI: 10.1039/cXsm00000x/}



\section{Introduction}

It may be surprising, but crystalline solids can always flow. This happens not only in defective, i.e. finite temperature, solids via the motion of dislocations \cite{seeger1955cxxxii,rice1974ductile}, but also in perfect crystals, where a finite load induces distortions to relax the stresses \cite{ruelle1999,lebowitz1968statistical}. However, the solid's viscosity diverges as the applied deformation vanishes \cite{sausset2010solids}, which is consistent with our everyday experience. 

Hence ,ideal crystals can undergo relaxation and observable flow if we are patient enough: yet, it is not clear how ordered solids enter the flowing state and how this process depends on the specific symmetries and composition of the crystal. Theoretical approaches based on nucleation theory \cite{sausset2010solids,nath2018} provide a promising route to account for the emergence of flow. Such approaches are however valid at vanishingly small stresses (or strains): a different scenario is expected for finite, large loads.

In an inspiring work by Sausset \textit{et al.} \cite{sausset2010solids} it was conjectured that (in analogy with ferromagnetic systems \cite{pyragas2000}) the application of oscillatory shear to crystalline solids would reveal a nonequilibrium,  dynamical phase transition between a non-flowing regime (elastic, reversible and ordered) and a flowing regime (plastic, irreversible and amorphous) triggered by nucleation at a particular shear amplitude or frequency. 

Here we precisely tackle such conjecture, and study a reversible-irreversible dynamical phase transition in several models under oscillatory shear. Depending on the dimensionality of the model and the nature of the ordered state, different order-disorder and reversible-irreversible transitions are observed as a function of the external driving amplitude. The dynamical transition that we observe is distinct from previous disorder-disorder transitions \cite{corte2008} and shear-melting studies \cite{stevens1991shear}. In particular, the transition we are ineterested in is an intrinsically nonequilibrium transition between different periodically driven steady states, as opposed to shear-melting transitions, where crystalline solids are linearly strained until their mechanical failure \cite{stevens1991shear,stevens1993simulations}.

In our analysis, we focus on the generic phenomenology of the transition, exploring how it depends on the dimensionality and the composition of the ordered phase. The transition is explored as a function of the amplitude of the periodic shear and the temperature of a stochastic thermostat coupled to the system. Given the generality of our investigation, we do not explore the effects of the particular orientation of the ordered state with respect to the shear direction or of the particular implementation of the thermostat. These details are known to affect the mechanical response  of crystals \cite{stevens1993simulations}, but do not influence the general features of the transition that we discuss.

 In our analysis we find that in two dimensions, for one-component systems whose ordered state is hexagonal, a continuous transition takes place. In three dimensions, a discontinuous transition from a flowing to a non-flowing steady state is observed, with a particular scaling for the relaxation time to steady state and the distribution of the amorphous region. suggesting the that an instability analogous to spindoal decomposition dominates the amorphization instead of nucleation only. 
 
The article is organised as follows: we first introduce the three studied models (2D single component Lennard Jones (LJ) system, 3D single component LJ and 3D two-component LJ), and specify the preparation and external driving protocols; we then sequentially discuss the phenomenology of the dynamical in transition for the three models, as identified by their specific order parameters; we finally focus on the 3D two-component LJ case, as it shows the clearest evidence for a dynamical transition driven by a spinodal instability; we summarise and discuss the implications of our observations in the conclusions.

\section{Models and methods}
\label{sec:models}
 We perform nonequilibrium molecular dynamics simulations of particle systems in two (2D) and three dimensions (3D) with Lennard Jones interactions. Each system is prepared in an ordered state at different temperatures and deformed according to a sinusoidal uniaxial shearing protocol of period $\tau_0$ and maximal strain amplitude $\gamma_0$. In two dimensions, we consider the LJ hexagonal lattice at fixed reduced density $\rho=0.86$, which is approximately $8\%$ larger than the solid density at the triple point \cite{barker1981}. In three dimensions, we consider two initial solid states: Firstly, the face centered cubic (fcc) LJ crystal, which is the stable phase for this model, at reduced density $\rho=1.0$, with melting temperature $T_{\rm melt}^{LJ3d}=0.96$ \cite{mastny2007}; Secondly, the $\rm Al_2 Cu$ crystal structure corresponding to the low energy crystalline state of a popular LJ nonadditive binary mixture of large (A) and small (B) particles \cite{Kob1994}, prepared at pressure $P=0$ with estimated melting temperature $T_{m}^{BLJ}=0.447(2)$ \cite{crowther2015}.


We study the amorphization of ordered phases in two (2D) and three (3D) dimensions through oscillatory shear. For the sake of simplicity, we focus on systems with spherically symmetric interactions which are initially prepared in their respective thermodynamically stable ordered phases.

The preparation protocol of the several systems is detailed in the following sub-sections.

\subsection{Two dimensional single component case} We consider a single component system of particles interacting via the Lennard-Jones potential 
\begin{equation}
V(r)=4\varepsilon \left[ (\sigma/r)^{12}-(\sigma/r)^6\right].
\label{seq:lj}
\end{equation} The parameters $\varepsilon$ and $\sigma$ provide the units for energy and length scales, the particle mass $m$ is set to unity as well as the Boltzmann constant $k_B$, defining a set of reduced units. The potential is truncated and shifted at the conventional cutoff distance $r_{\rm cut}=2.5\sigma$ to speed-up the calculations. 

For sufficiently high densities and low temperatures, the system attains an ordered hexagonal phase \cite{gribova2011}. This is a phase with only quasi-long range order, i.e. where spatial correlations decay slowly (algebraically as opposed to exponentially). We directly arrange the particles on an hexagonal lattice at number density $\rho = 0.86$, randomly initialize the velocities and equilibrate the ordered phase in the isothermal-isochoric (NVT) ensemble employing a deterministic Nose-Hoover thermostat with damping time $t_{\rm damp}=0.1$ reduced time units in a range of temperatures $T\in[0.4, 0.6]$.  We consider systems of sizes $N=2800,5000$ particles.

In order to characterise the degree of local order, we measure the global hexagonal order parameter $\Psi=|1/N \sum_k \psi_k|$, with 

\begin{equation}
	\psi_k = N_{\rm neig}^{-1}\sum_{l \in \{ \rm neig\}_k }\exp[6i\phi_{kl}]
\end{equation}
where $\{\rm neig\}_k$ is the list of the $N_{\rm neig}$ neighbours of particle $k$, $\phi_{kl}$ is the orientation of a vector connecting the centres of particle $k$ and the neighbour $l$ with respect to the a reference axis, and $i$ is the imaginary unit.

\subsection{Three dimensions} For the three-dimensional case we consider two different ordered states: a single-component LJ crystal and two-component LJ crystal.

	 \subsubsection{Single-component case}
	 
	 We employ precisely the same interaction potential as in Eq.\ \ref{seq:lj} which in 3D leads to the formation of a genuine crystalline face-centered-cubic (fcc) solid phase at sufficiently high densities and low temperature \cite{mastny2007melting}. Here we report results for a particular reduced density $\rho=1.0$ and a range of temperatures $T\in[0.6,1.4]$. Notice that so-called quasi-universality in the Lennard-Jones systems ensures that these specific choices are not particularly important for the overall phenomenology of the dynamical transition under study \cite{bacher2014explaining}. 
	 
	 Similar to the 2D case, the system is prepared in an ordered state through equilibration in the NVT ensemble via a Nos\'e-Hoover thermostat with damping time $t_{\rm damp}=0.1$ reduced units. We consider systems of $N=4000$ particles. 
	 
	 We track structural changes using the ring-statistics \cite{stukowski2012} and Voronoi-based metrics \cite{Malins2013a,lazar2017} to measure face centered cubic (fcc), body centered cubic (bcc) and hexagonal close packed (hcp) arrangements.
	 
	\subsubsection{Two-component case}
	
	We also consider a binary crystal with interaction parameters originally fixed by Kob and Andersen to model the behaviour of a metallic glassformer \cite{Kob1994}. Here we have large (A) and small (B) particles interacting via Lennard-Jones potentials tuned to to favour the A-B interaction, so that $\varepsilon_{BB}=0.5\varepsilon_{AA}$, $\sigma_{BB}=0.88\sigma_{AA}$ and the cross interaction parameters are $\varepsilon_{AB}=1.5\varepsilon_{AA}$, $\sigma_{AB}=0.8\sigma_{AA}$. All quantities are expressed in terms of the large particles size $\sigma_{AA}=\sigma_A=\sigma$ and energy scale $\varepsilon_{AA}=\varepsilon$ and mass $m_{AA}=m_{BB}=m$. In the 2A:1B composition the system has been shown to grow at zero reduced pressure $p=0$ the $\rm Al_{2}Cu$ crystalline lattice \cite{crowther2015}. For consistency, we prepare our ordered samples at zero pressure and  consider systems of several sizes $N=768,1500,12000, 96000, 768000$. To identify the corresponding density, we construct the equation of state following Vinet's law for Lennard-Jones crystals
\begin{equation}
p = 2 B_0 \, x^3 (x^2 - 1),
\label{eq:vinet}
\end{equation}
where $x=\rho/\rho_0$ and $B_0$ and $\rho_0$ are the bulk modulus and density at zero pressure.
  
As shown in the Supplementary Information$^\dag$ this expression provides a very good fit to the molecular dynamics results. It also allows us to estimate the temperature dependence of the bulk modulus $B_0$ and of the density at zero pressure $\rho_0$. A linear fit to low temperatures provides an estimate for the density at zero pressure and temperature $\rho_0(T=0) \approx 1.48$ in agreement with previous estimates \cite{Fernandez:2003he}. 

For the binary crystal $\rm Al_2 Cu$, we use ring-statistics \cite{stukowski2012} and Voronoi-based metrics \cite{Malins2013a,lazar2017} to measure the number of B particles in the center of a bicapped square antiprism (simply \textit{antiprism} in the following), which is the distinctive motif of the crystal unit cell and a recurrent local motif (or \textit{locally favourite structure}) in the melt \cite{coslovich2007,malins2013faraday}.

\subsection{Oscillatory shear protocol}

For our nonequilibrium protocol, we shear the several systems according to a sinusoidal deformation as implemented in the molecular dynamics package LAMMPS \cite{plimpton1995fast}. The deformation is uniaxial along the x direction so that the strain as a function of time oscillates as in
\begin{equation}
	\gamma(t) = \gamma_0 \sin(2\pi t/\tau_0),
	\label{eq:protocol}
\end{equation}
where $\gamma_0$ is the maximum shear amplitude and $\tau_0$ is the period of the oscillations. The obtained triclinic box is periodic in all dimensions (effectively implementing Lees-Edwards's boundary conditions).

During the deformation, the temperature of the system is kept constant by a Langevin thermostat whose characteristic damping time is $t_{\rm damp}=100dt$, where $dt$ is the time step for the integration of the equations of motion, set to $dt=0.001$. For the different Lennard-Jones models considered here, we keep the period of the oscillations constant to $\tau_0=10$ in reduced Lennard Jones units and vary the other physical parameters, such as the amplitude of the oscillations or the temperature.

We typically consider 10 distinct realizations of the dynamics for a given set of parameters, starting from different initial conditions for both the prepared ordered state and the particle velocities, initialized from a pseudo-random sample of the Maxwell-Boltzmann distribution.
\section{Results on the order-disorder dynamical transition} We are interested in the transition between two distinct nonequilibrium steady states (NESS): an ordered NESS, where the system remains in the prepared ordered state, and an amorphous (or only semi-crystalline) NESS, attained while the system is continuously subject to oscillatory shear and to modest thermal fluctuations. 

We start with the analysis of the 2D LJ system; continue with the 3D LJ single component system and finally focus on the 3D two component (binary) LJ system.

\subsection{Single-component 2D Lennard-Jones hexagonal solid}

\begin{figure}[t]
  \centering
  \includegraphics{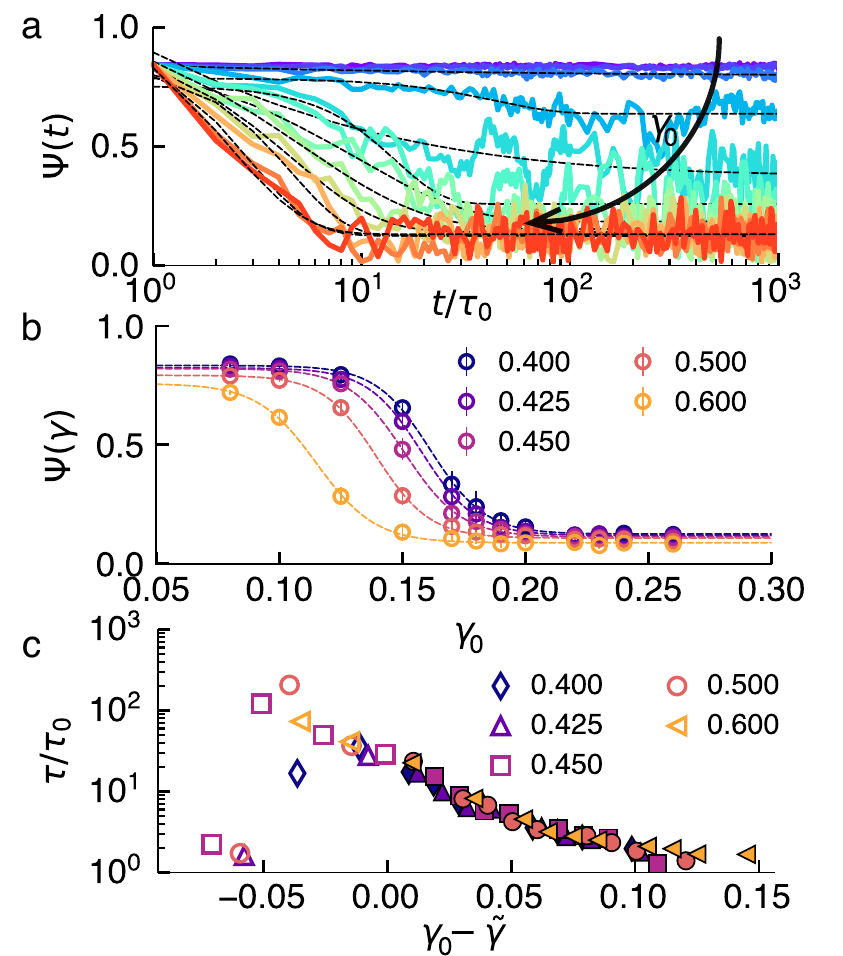}
  \caption{\textbf{Dynamical order-disorder crossover in the 2D hexagonal lattice.} (a) Evolution of the single-run global hexagonal order parameter $\Psi$ for $\gamma_0\in[0.5,0.26]$. Dotted black lines are stretched exponential fits. (b) Steady state $\Psi_{\infty}$ for different temperatures, averaged over several trajectories. (c) Relaxation time to the steady state $\tau$ in units of the oscillation period $\tau_0$ as a function of the distance from the crossover strain amplitude $\tilde{\gamma}$. Full relaxation within the time of the simulations is for $\gamma_0>\tilde{\gamma}$ (filled symbols).}
  \label{figLJ2d}
\end{figure}

Two-dimensional LJ disks, in absence of external driving, undergo a liquid-hexagonal transition which proceeds via a Kosterlitz-Thouless-Halperin-Nelson-Young mechanism, i.e. through a so-called hexatic phase \cite{Wierschem2011,gribova2011} in a weakly first-order phase transition. The equilibrium ordered phase is known to be characterised by a mosaic of locally ordered (hexagonal) patches  and amorphous structures \cite{patashinski2010}. Here we promote the amoprhization of the hexagonal ordered state through sinusoidal shear of variable strain amplitude at several temperatures, $T\in[0.4,0.6]$. Starting from a thermalised hexagonal state, we monitor the time evolution of the global orientation order parameter $\Psi$ for different values of the maximum strain amplitude $\gamma_0$, see Fig.~\ref{figLJ2d}(a). For small values of $\gamma_0$, the system responds elastically, and at every cycle it reversibly attains the initial ordered state characterised by large values of $\Psi$. However, as we increase the maximum amplitude, we observe that the orientational order parameter gradually relaxes to lower and lower values. We fit the relaxation curves with a stretched exponential $\Psi(t)=\Psi_{\infty}+(\Psi_0-\Psi_\infty)\exp[-(t/\tau)^b]$, from which we derive the relaxation time $\tau$, the long and short time value of the order parameter $\Psi_\infty$ and $\Psi_0$ and the stretching exponent $b$ which tends to unity from below as we approach larger strain amplitudes. 

As shown in Fig.~\ref{figLJ2d}(b), the transition from an ordered to a disordered steady state occurs over a broad range of strain amplitudes  $\gamma_0$ at all the considered temperatures, so that we can successfully fit it with a hyperbolic tangent function $\Psi_{\infty}(\gamma_0)=a+b\tanh[(\gamma_0-\tilde{\gamma})/c]$, where $\tilde{\gamma}$ is the fitting parameter corresponding to the inflection point of the curve.
Plotting the resulting relaxation times $\tau(\gamma_0)$ as a function of the distance from the inflection point $\tilde{\gamma}$ we manage to collapse the data from different temperatures on a semilogarithmic plot, Fig.~\ref{figLJ2d}(c), suggesting that the temperature dependence is mainly encoded in the crossover value $\tilde{\gamma}$. Only for $\gamma<\tilde{\gamma}$ the collapse fails, but this is because reliable values of the relaxation time cannot be obtained through the stretched exponential fit when the system relaxes on time scales larger than the simulation time scale. For $\gamma>\tilde{\gamma}$, we observe that in the vicinity of $\tilde{\gamma}$, the relaxation time increases by less than two orders of magnitude with respect to the large $\gamma_0$ regime.

The transition can be analysed in terms of its effect on local orientational order, in terms of the the complex number $\psi_i$ associated to every particle $i$. The spatial distributions of its modulus and argument allow us to keep track of the formation of boundaries between differently oriented regions.

\begin{figure}[t]
  \centering
  \includegraphics[width=\columnwidth]{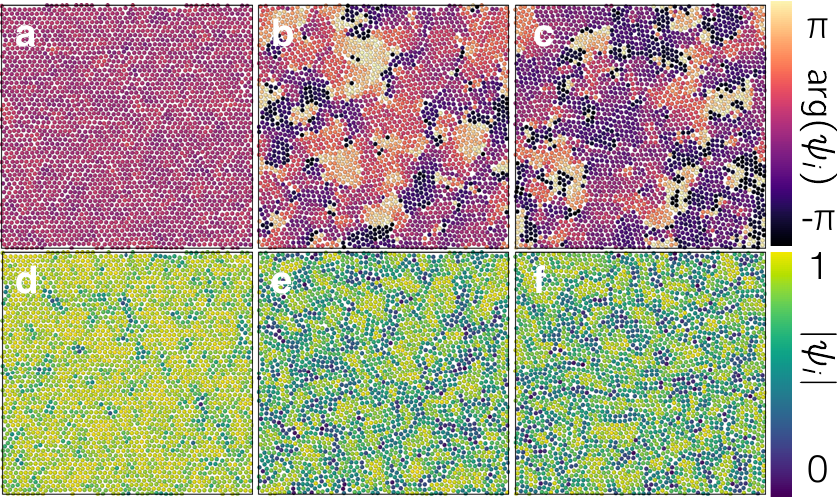}
  \caption{\textbf{Dissolution of the local hexagonal order for two-dimensional system.}  Color-coded configurations at temperature $T=0.40$ and strain amplitude $\gamma_0=0.22$. Top row: local order as detected by the angular part of the local orientational order parameter $\psi_i$. Bottom row: the same, as detected by the modulus of $\psi_i$. Columns (a,d), (b,e) and (c,f) correspond to identical time frames $t=0, 50,100\tau_0$.}
  \label{fig:2dorder}
\end{figure}

In Fig.\ \ref{fig:2dorder} we present configurations color-coded by the values of the argument and the modulus of $\psi_i$ respectively, for a particular temperature $T=0.4$ and strain amplitude $\gamma_0=0.22$ over several periods of oscillatory shear.

Both the modulus and the argument of $\psi_i$ identify domains of coherence and local order. However, grain boundaries lacking 6-fold coordination are more immediately identified in the modulus, as low values of $|\psi_i|$ typically correspond to four-fold coordinated particles.

The reduction of quasi-long range order over time takes place through the formation of extended defect lines which break the lattice into sub-domains of smaller sizes. This demonstrates that orientational correlations span over smaller scales as the system reaches the disordered steady state. The size of the orientational coherence can be estimated from the decay of the conventional two-point correlation function
\begin{equation}
	g_6(r)=\langle \psi^*({\bf r}) \psi({\bf r})\rangle,
\end{equation}
where the average here is only performed on all the particles at a given time $t$.

\begin{figure*}[t]
  \centering
  \includegraphics{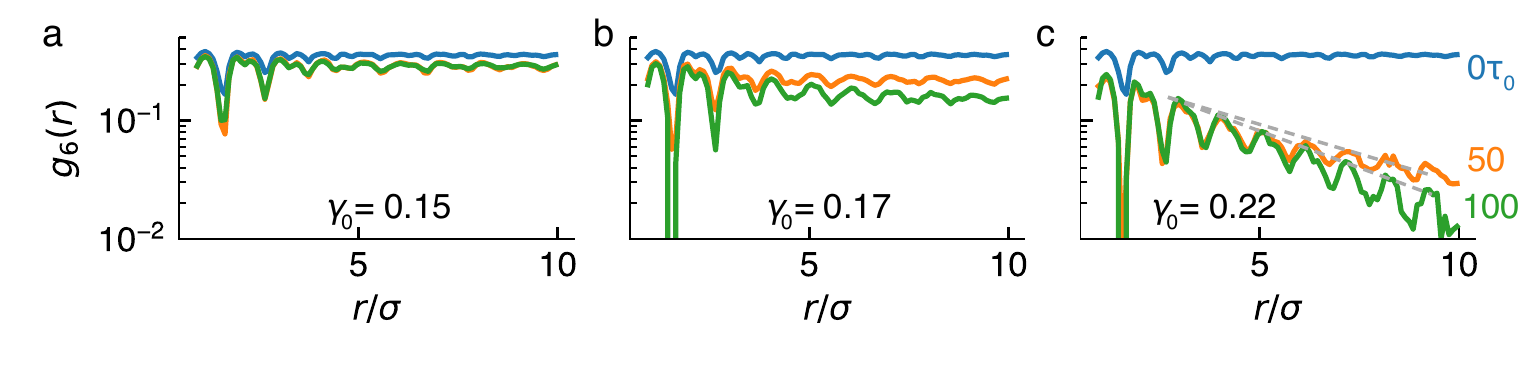}
  \caption{\textbf{Orientational pair correlation function in the 2D LJ system}. At temperature $T=0.40$ for three distinct values of the strain amplitude $\gamma_0=0.15,0.17,0.22$ across the order-disorder crossover, the orientational correlations as measured by $	g_6(r)=\langle \psi^*({\bf r}) \psi({\bf r})\rangle$ decay very mildly for small strain amplitudes, and on much shorter scale for large amplitudes. In every, correlations are plotted after $0,50,100$ shear cycles. Dashed lines are exponential fits to extract the correlation length $\xi$. }
  \label{fig:g6}
\end{figure*}

In Figure \ref{fig:g6} we plot $g_6(r)$ across the order-disorder crossover at a relatively cold temperature $T=0.4$ for after 0, 50 and 100 shearing cycles of duration $\tau_0$. For $\gamma_0=0.15$ correlations are strong well beyond ten particle diameters. In fact, the decay of the correlations can be fitted equally weel by exponential or power law decays, indications that correlations are quasi-long range, as in the hexatic phase of equilibrium disks. However, across the transition, orientational correlations decay more steeply and undergo a time evolution which is more rapid as the strain amplitude increases. At large strain amplitudes, $\gamma_0=0.22$, orientational correlations decay over a length $\xi(\gamma_0=0.22)\approx5\sigma$.

Hence, in two dimensions, we observe a dynamical \textit{continuous} crossover taking place through the formation of large regions of different orientations, which become progressively smaller as the strain amplitude is increased. Notice that at the chosen temperature and densities the quiescent (i.e. not driven) system would simply form an ordered hexagonal phase. The driven system, instead, present a wide spectrum of possible steady states of increasing orientational order as the amplitude of the strain oscillations is reduced.

\subsection{Single-component 3D Lennard-Jones crystal}
\begin{figure}[tp]
  \centering
  \includegraphics{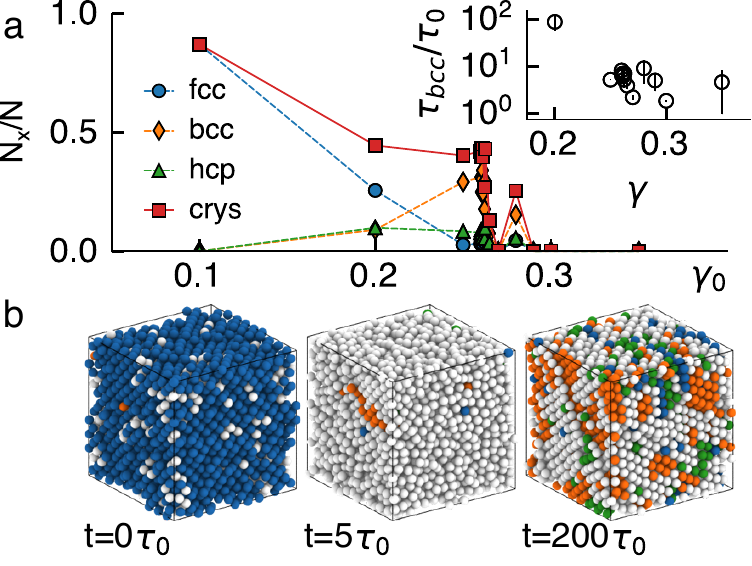}
  \caption{\textbf{Dynamical order-disorder transition in the 3d single-component fcc crystal.} (a) Fraction of particles in crystalline arrangements (fcc, bcc, hcp and the sum crys=fcc+hcp+bcc) as a function of $\gamma_0$. Inset: bcc nucleation time $\tau_{bcc}$ with respect to $\gamma_0$. (b) Time evolution of the structure of the sheared crystal at $\gamma_0=0.28$ at time $t=0,5, 200 \tau_0$, color-coded as in (a).}
  \label{figLJ3d}
\end{figure}

A very different scenario emerges from the three dimensional simulations. We first consider the single-component fcc crystal of LJ particles sheared along the $\langle100\rangle$ orientation. We follow a protocol analogous to the two-dimensional case, but now we monitor (via Common Neighbour Analysis \cite{stukowski2012,stukowski2009} and Voronoi cell analysis as implemented in VoroTop \cite{lazar2017}) a variety of multiple crystalline motifs: fcc, hcp, bcc. We prepare an equilibrated fcc crystal at reduced density $\rho=1.0$ at temperatures $T\in [0.6,1.4]$, from around the triple point to well above the melting temperature. While at high temperature bulk melting dominates, interesting transitions emerge at low temperatures. In Fig.~\ref{figLJ3d} we present the case of $T=0.6$.

As we increase the shear amplitude $\gamma_0$ the crystalline cells become highly distorted, so that a perfect fcc crystal is transformed into a defective fcc/hcp crystal with bcc inclusions, see Fig.~\ref{figLJ3d}(a) and \dag. However, for $\gamma_0\sim0.26$ (corresponding to the constrained tension ideal strain of the fcc LJ crystal of $\gamma_{\rm ideal}=0.26$ \cite{macmillan1972}) an abrupt transition to a non-crystalline steady state is observed. Interestingly, the structure of the crystal for $\gamma_0\rightarrow \gamma_{\rm ideal}$ is dominated by the bcc order instead of the fcc order. This is somehow surprising: for the quiescent bulk, the free energies of the hcp and fcc structures are close (with the fcc crystal being marginally lower in free energy) while the bcc crystal is unstable \cite{travesset2014phase}. However, it has been suggested that precritical nuclei are formed in the bcc arrangement \cite{rein1996numerical}: in our case, the oscillatory protocol appears to allow such nuclei to grow and become stable at steady state.

The transformation from fcc to bcc for $\gamma_0<\gamma_{\rm ideal}$ occurs with very limited amorphization: the rearrangement of the crystalline defects is sufficient to cause the change in crystalline symmetry. However, for larger $\gamma_0$ and in particular for $\gamma_0=0.28$ we observe partial re-crystallization of the system: as depicted in Fig.\ref{figLJ3d}(b) the initial fcc crystal is initially turned into a largely amorphous packing in a few cycles. However, bcc crystal nucleation and growth occurs in the following cycles so that the system attains a largely re-crystallized NESS. Hence, amorphization and bcc-nucleation are competing processes for $\gamma_0\sim \gamma_{\rm ideal}$. In such a regime, the time to attain an amorphous steady state $\tau$ is longer than the nucleation time for the bcc crystal $\tau_{\rm bcc}$: this means that the fcc-order to disorder transition is mediated by an intermediate family of nonequilibrium bcc steady states. 

\subsection{Two-component Lennard-Jones crystal}
The results on the single-component Lennard-Jones crystal motivate us to consider a system less prone to re-crystallization. Previous studies on glass-forming systems have shown that a non-additive 2:1 mixture of large and small Lennard-Jones particles is highly stable to crystallization \cite{crowther2015}, probably due, firstly, to the vicinity of its stoichiometry to the eutectic point of the mixture \cite{pedersen2018} and secondly to the fact that binary crystals fail differently under shear compared to one-component crystals \cite{horn2014does}. The mixture crystallizes into the $\rm Al_2 Cu$ arrangement, whose unit cell is characterized by spindles of small particles surrounded by square rings of large particles, forming a \textit{bicapped square antiprism} \cite{fernandez2003}. In fact, this is the same symmetry of recurrent geometrical motifs (also called \textit{locally favoured structures}) in the liquid and supercooled liquid states of the same model, which play an important role in the emergence of slow relaxation within the glassy state \cite{coslovich2007,malins2013faraday, turci2017}. It has been shown that the number of particles involved in such motifs can be large in the liquid at low temperature without any sign of spontaneous crystal formation  \cite{crowther2015}. We therefore expect that a large enough deformation of an equilibrated configuration in the $\rm Al_2 Cu$ crystalline arrangement of the model would lead to a substantial amorphization of the material which would be irreversible, with no secondary nucleation processes.

\begin{figure*}[ht]
	\centering
	\includegraphics{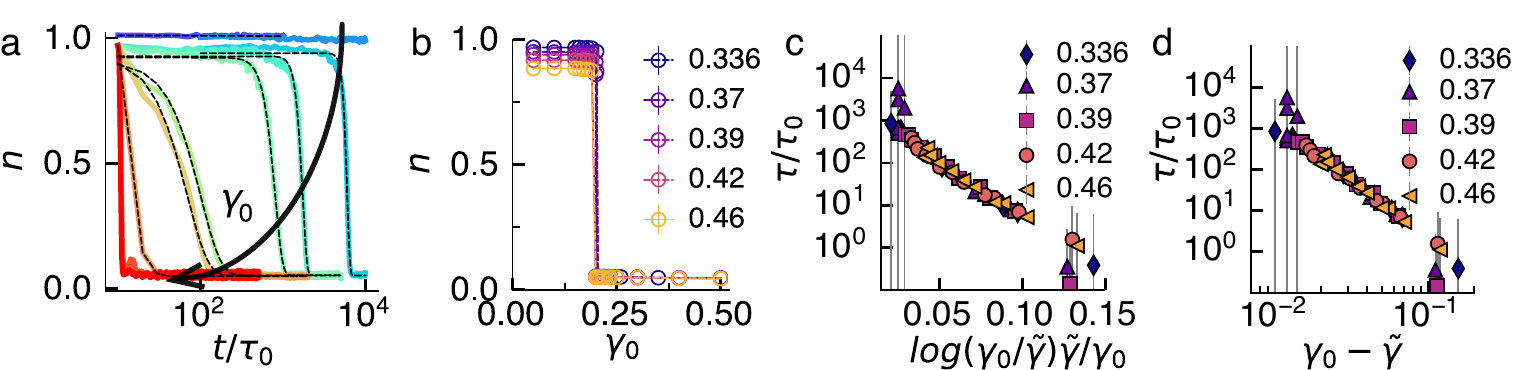}
	\caption{\textbf{Dynamical order-disorder transition in the 3d binary crystal.} (a) Fraction of particles $n(t)$ forming bicapped square antiprisms for $\gamma_0\in[0.19, 0.5] $ at $T=0.336$. Dashed lines are stretched exponential fits, defining the relaxation time $\tau$. (b) Steady-state values of $n$ for different temperatures: as a function of $\gamma_0$ we observe a sharp transition around $\tilde{\gamma}\approx0.19$ between an ordered and disordered dynamical phase. In (c) and (d) we present two empirical scalings of the relaxation time $\tau$: (c) a collapse with respect to $\log(x)/x$ with $x=\gamma_0/\tilde{\gamma}$ and a critical-like scaling $\tau\propto (\gamma_0-\tilde{\gamma})^{-\nu}$, from which $\nu\approx3.0$. }
	\label{figBLJ}
\end{figure*}

We monitor for this purpose the specific crystalline order of the binary crystal employing Voronoi face analysis (in which the antiprism can be detected with the [0,2,8] signature) as well as the Topological Cluster Classification method \cite{Malins2013a}. We have seen that in the two-dimensional case the increase of disorder is  gradual in time, while in the three-dimensional one-component LJ this is mediated by the nucleation of a bcc phase within the original fcc phase. In the case of the BLJ crystal (for a choice of period $\tau_0=10$ in Lennard-Jones time units with respect to the large particles) we directly observe the formation of disordered pockets, without the interference of intermediate crystal formation. The time evolution of the fraction of particles in antiprisms, Fig.~\ref{figBLJ}(a), shows that when the shear amplitude is small ($\gamma_0\lesssim 0.19$), the system remains ordered for several hundreds of cycles before rapidly and irreversibly falling into the amorphous state. 

In Fig.~\ref{figBLJ}(b) we plot the steady state value of the fraction of particles in an antiprism for several values of the temperature. As expected, the transition from a non-flowing to a flowing regime is very sharp with no further crystallisation beyond a characteristic value of $\tilde{\gamma}$. We also emphasize that higher temperatures correspond to slightly smaller steady state populations of antiprisms, both in the non-flowing and in the flowing dynamical phase.

As suggested by Fig.~\ref{figBLJ}(a) the relaxation time $\tau$ at which the irreversible plastic event occurs increases as $\gamma_0\rightarrow\tilde{\gamma}^+$. In Fig.~\ref{figBLJ}(c) and (d) we show two different scalings of this relaxation times: inspired by the estimate $\log \tau\propto \log(x)^\delta/x$ with $x=\gamma_0/c$ proposed in \cite{sausset2010solids} for the nucleation of amorphous patches in the limit of small stresses (where $c$ is a material specific constant and $\delta=3$ in the original calculations in three dimensions) we explored a $\log(x)/x$ scaling finding that only $\delta=1$ would account for our data, Fig.~\ref{figBLJ}(c). However, if for $\gamma<\tilde{\gamma}$ the steady states are truly irreversible, then $\lim_{\gamma_0\rightarrow\tilde{\gamma}^+}\tau=+\infty$, so we also model the singularity at $\tilde{\gamma}$ with a critical power law scaling, obtaining an equally good collapse of the data, see Fig.~\ref{figBLJ}(d), with $\tau\propto(\gamma_0-\tilde{\gamma})^{-\nu}$ with $\nu\approx3.0$.

\section{Analogy to spinodal decomposition in the 3D binary LJ crystal}

The 3D binary crystal presents a clear transition at a specific value of the strain amplitude and the timescales appear to diverge in a fashion compatible with scale-free, power law scaling. This suggests the possibility that the transition is dominated by a barrier-less process akin to spinodal decomposition, with the timescales for the transition from the reversible solid NESS to the irreversible amorphous NESS being controlled by the coarseing of the amorphous regions.

\begin{figure}[t]
	\centering
	\includegraphics{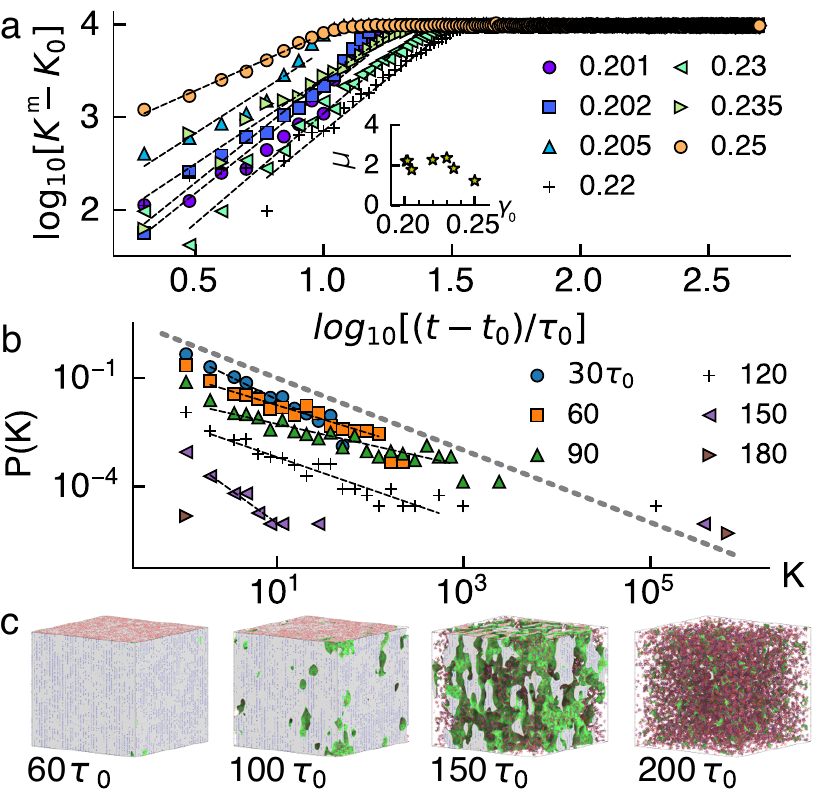}
	\caption{\textbf{Growth of amorphous regions in the 3D binary crystal.} (a) Number of particles in the largest amorphous cluster $K^{\rm m}$ for a BLJ crystal of $N=12000$ particles at $T=0.336$ for several $\gamma_0$. Time $t_0$ is chosen so that $K_0=K^m(t_0)=100$ particles. Dashed lines are power-law fits $K^{\rm m}(t)\propto t^{\mu(\gamma_0)}$ and in the inset we report $\mu(\gamma_0)$. (b-c) Time evolution of a system of $N=768000$ particles at $\gamma_0=0.215$ and $T=0.336$: (b) probability distribution of the sizes of amorphous patches, with the $1/K$ slope (dashed line); (c) successive snapshots of the system,  with crystal/amorphous interfaces (green surfaces, grey the periodic faces) dotted by antiprism particles (blue and red).}
	\label{figlengths}
\end{figure}

To test this, we further investigate the scaling behaviour measuring the growth of amorphous patches within the crystal in large simulations of up to $N=768000$ particles. We identify connected regions without antiprisms and measure the number of particles  $K$ in each cluster. In Fig.~\ref{figlengths} (a) we show how the largest value $K^{\rm m}$ increase as time evolves for different $\gamma_0$ observing that $K^{m}\propto t^{\mu}$ before reaching a plateau value once the system turns amorphous, with the exponent reaching  $\mu\approx 2$ for the smallest $\gamma_0$ for which we still observe amorphization. If we assume that the cluster length is approximately $\ell=\sqrt[3]{K}$ we have $\ell\propto t^{1/z}$ with the dynamical exponent $z=3/\mu\approx3/2$ to be compared with $z=2$ and $z=3$ of Model A (e.g. Ising) and model B (e.g. Cahn-Hilliard model for spinodal decomposition) respectively \cite{hohenberg1977theory}. At the same time, the probability distribution of the cluster sizes $P(K)$ appears to be inherently scale-free, as shown in  Fig.~\ref{figlengths} (b) for the case of $\gamma_0=0.21$ at $T=0.336$, where, for the largest accessible system, we measure $P(K)\propto1/K$ until one full relaxation time. The spatial  distribution of clusters corresponds to a branched network of amorphous regions across the system, as pictorially illustrated in the snapshots in Fig.~\ref{figlengths} (c), where the amorphous domains appear to span across the length of the system. The growth pattern of the amorphous network is reminiscent of coarsening or spinodal decomposition, occurring in other protocols of fast energy injection, such as superheating phenomena \cite{belonoshko2007}.

As a second test for a spinodal-like scenario, we consider the seeding of amorphous regions within the solid. In analogy to studies of crystal nucleation in metastable liquids \cite{espinosa2016seeding}, we probe whether the sharp transition in the binary Lennard-Jones mixture follows a nucleation/growth scenario through the seeding of disordered regions within the crystalline state.

We consider a system of $N=12000$ particles at low temperature $T=0.336$ and a strain amplitude of $\gamma_0=0.21$,  10\% larger than the estimated critical strain $\tilde{\gamma}\approx0.19$. In such conditions, we know that the relaxation from the crystal to the liquid occurs on timescales of the order of $10^3 \tau_0$ (see Fig.\ 3(d) in the main text).

\begin{figure*}[bt]
  \centering
  \includegraphics{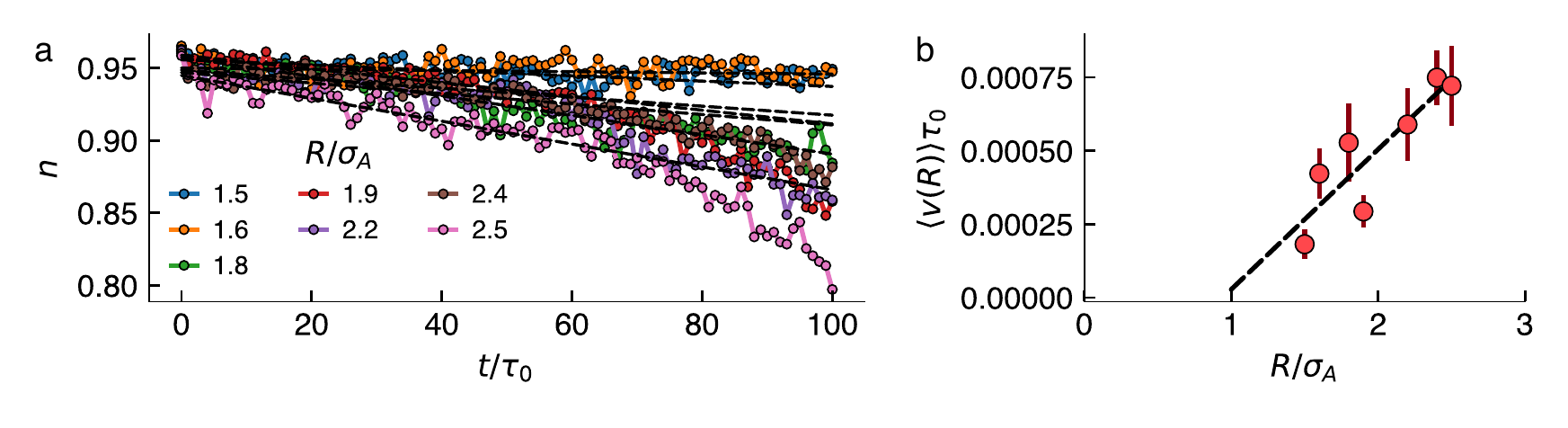}
  \caption{\textbf{Growth of seeded amorphous regions in the BLJ crystal.} (a) Time evolution of the fraction of particles in crystalline regions for different values of the radius $R$ of the seeded amorphous nucleus, at temperature $T=0.336$ and strain amplitude $\gamma_0=0.21$. Dashed lines are linear fits to the initial decay, defining the growth speed $v=dn/dt|_{t=0}$  (b) Average growth speed $\langle v(r)\rangle$ as a function of the radius of the seeded amorphous nucleus: a linear fit points to zero growth speed only for nuclei of size $R\approx1\sigma_A$. See video in the Electronic Supplementary Material$^\dag$.}
  \label{fig:seed}
\end{figure*}
We first prepare a crystalline arrangement of particles through the equilibration procedure described in Sec.\ \ref{sec:models}. We then identify particles outside a given radius $R$ from the origin of the reference frame and freeze their positions. For the particles within the radius $R$ we reassign their velocities, randomly selected them from a Maxwell-Boltzmann distribution, so that their kinetic temperature is $T_{\rm seed}=10$ and solve the equation  fo motion (with frozen external particles and with no shear) for a time of approximately 100$\tau_0$. At such high temperatures the particles can cross each other more easily and the spherical region becomes disordered. Following this route we prepare 10-30 initial configurations from which we start oscillatory shear simulations. We track the number of particles in crystalline motifs $N_x$ and the relative fraction $n=N_x/N$ over time and obtain single-run curves as in Fig.\ \ref{fig:seed}(a). We define the growth speed 
\begin{equation}
	v = \left.\frac{dn}{dt}\right|_{t=0}
\end{equation}
and numerically evaluate it through a linear fit at early times. Averaging over several realizations, we obtain $\langle v(R)\rangle$ as a function of the radius of the disordered seed, see Fig.\ \ref{fig:seed}(b). We observe that we only measure positive growth speeds and that a linear fit through the data points $\langle v(R)\rangle=0$  only for $R\approx1$. Such a small size for the initial seed can hardly be called a nucleus. See video in the Supplementary Information $\dag$ for an example trajectory.

From the seeded nucleation analysis we infer that amorphous nuclei of very small sizes appear to lead to the irreversible transition from the ordered to the amorphous oscillatory steady state. We interpret this as a further indication in support of a spinodal-like instability scenario.

\section{Conclusions}

Under the action of an oscillatory external driving, atomistic models of ordered states in two and three dimensions undergo a novel nonequilibrium dynamical transition between a non-flowing, ordered phase and a flowing amorphous phase. This -- originally only conjectured -- dynamical transition is revealed to have a very rich phenomenology: in 2D, it appears  merely as a broad crossover between a disordered state with residual orientational order and an ordered hexagonal phase; in 3D, it corresponds to a sharp transition that can be masked by the nucleation of ordered phases with symmetries different from the ones of the unperturbed solid (i.e. bcc instead of fcc); moreover, with a suitable choice of the ordered phase, as in the case of the binary Lennard-Jones crystal, the 3D transition occurs abruptly in a narrow range of shear amplitudes with a scaling of the time required to relax to the amorphous state which is compatible with power-law, critical scaling. 

The sharpest transition is therefore observed in the case of the two-component system with non-additive Lennard-Jones interactions, which prevent recrystallization across the dynamical transition. In the case of this particular system, the numerical results suggest that -- even for small systems -- one can define a critical strain amplitude $\gamma_0$ at which amorphous pockets are formed and start growing irreversibly, in a process akin to coarsening. The relatively high density of amorphous pockets, their structure and growth are reminiscent of spinodal decomposition as opposed to a nucleation scenario, which is confirmed by the study of the spontaneous growth of very small seeded amorphous regions within the ordered phase. Coarsening of the emergent disordered phase appears to play a role as important as the  nucleation-based description described in Ref.\cite{sausset2010solids}, which dominates the onset of flow in the small strain limit.

The phenomenology presented here can be tested in experiments, for example with colloidal crystals in shear cells, both in two and three dimensions. Further theoretical study will focus on the possible implications of the present dynamical transitions for amorphous solids such as glasses and gels, where critical behaviour is believed to play an important role in the emergence of rigidity \cite{jaiswal2016mechanical,parisi2017shear,ozawa2018random}.
\\

The authors would like to thank T.\ Speck, J.\ Lam, C.\ Valeriani, N.\ Wilding and R.\ L.\ Jack for insightful discussions. This work was carried out using the computational facilities of the Advanced Computing Research Centre, University of Bristol. FT acknowledges the support of the European Research Council under the FP7/ERC Grant agreement n$^{\circ}$617266 NANOPRS.	



\balance


\bibliographystyle{rsc} 

\providecommand*{\mcitethebibliography}{\thebibliography}
\csname @ifundefined\endcsname{endmcitethebibliography}
{\let\endmcitethebibliography\endthebibliography}{}

\end{document}